\documentclass[twocolumn]{aastex631}

\usepackage{url}

\usepackage{amsmath}
\usepackage{float}
\usepackage{graphicx}
\usepackage{multirow}
\usepackage{soul}
\usepackage{natbib}

\usepackage{rotating}
\usepackage{hyperref}

\newcommand{\bjdtdb}{\ensuremath{\rm {BJD_{TDB}}}}
\newcommand{\feh}{\ensuremath{\left[{\rm Fe}/{\rm H}\right]}}

\newcommand{\teff}{\ensuremath{T_{\rm eff}}}
\newcommand{\teq}{\ensuremath{T_{\rm eq}}}

\newcommand{\logg}{\ensuremath{\log g_*}}

\newcommand{\ecosw}{\ensuremath{e\cos{\omega_{*}}}}
\newcommand{\esinw}{\ensuremath{e\sin{\omega_{*}}}}
\newcommand{\secosw}{\ensuremath{\sqrt{e}\cos{\omega_{*}}}}
\newcommand{\sesinw}{\ensuremath{\sqrt{e}\sin{\omega_{*}}}}
\newcommand{\msun}{\ensuremath{\,M_\Sun}}
\newcommand{\rsun}{\ensuremath{\,R_\Sun}}
\newcommand{\lsun}{\ensuremath{\,L_\Sun}}

\newcommand{\mj}{\ensuremath{\,M_{\rm J}}}

\newcommand{\mplanet}{\ensuremath{\,M_{\rm P}}}
\newcommand{\mratio}{\ensuremath{\,M_{\rm P}/M_{*}}}

\newcommand{\rj}{\ensuremath{\,R_{\rm J}}}
\newcommand{\re}{\ensuremath{\,R_{\rm \Earth}}\xspace}
\newcommand{\me}{\ensuremath{\,M_{\rm \Earth}}\xspace}
\newcommand{\fave}{\langle F \rangle}
\newcommand{\fluxcgs}{10$^9$ erg s$^{-1}$ cm$^{-2}$}

\newcommand{\tess}{{\it TESS}}

\newcommand{\kms}{\,km\,s$^{-1}$}

\newcommand{\mstar}{\ensuremath{M_{*}}}
\newcommand{\rstar}{\ensuremath{R_{*}}}

\newcommand{\vsini}{\ensuremath{v\sin{i_*}}}

\newcommand{\vbeta}{\ensuremath{v_{\beta}}}
\newcommand{\vgamma}{\ensuremath{v_{\gamma}}}
\newcommand{\vzeta}{\ensuremath{v_{\zeta}}}
\newcommand{\vvline}{\ensuremath{v_{\rm line}}}

\newcommand{\degrees}{\ensuremath{^{\circ}}}

\newcommand{\siglevel}{3.7}
\newcommand{\hotboundary}{$0.0021^{+0.0006}_{-0.0008}$}
\newcommand{\coolboundary}{$0.00027\pm0.00002$}
\newcommand{\hotboundarywithourerr}{$2\times10^{-3}$}
\newcommand{\coolboundarywithourerr}{$3\times10^{-4}$}
\newcommand{\finallambda}{${41.8}^{+2.1}_{-2.0}$\degrees}

\newcommand{\nall}{135}
\newcommand{\nbd}{7}
\newcommand{\nhot}{64}
\newcommand{\ncool}{71}
\newcommand{\nhotlowmassratio}{50}
\newcommand{\nhothighmassratio}{14}
\newcommand{\ncoollowmassratio}{31}
\newcommand{\ncoolhighmassratio}{40}

\usepackage{framed}

\newenvironment{keyresult}{%
  \begin{framed}\noindent\textbf{}%
}{%
  \end{framed}
}

\begin{document}
\title{From Misaligned Sub-Saturns to Aligned Brown Dwarfs:\\ The Highest $M_{\rm p}/M{_*}$ Systems Exhibit Low Obliquities, Even around Hot Stars}

\author[0009-0007-3168-5072]{Jace Rusznak} 
\affiliation{Department of Astronomy, Indiana University, 727 East 3rd Street, Bloomington, IN 47405-7105, USA}
\author[0000-0002-0376-6365]{Xian-Yu Wang} 
\affiliation{Department of Astronomy, Indiana University, 727 East 3rd Street, Bloomington, IN 47405-7105, USA}
\author[0000-0002-7670-670X]{Malena Rice}
\affiliation{Department of Astronomy, Yale University, 219 Prospect St., New Haven, CT 06511, USA}
\author[0000-0002-0376-6365]{Songhu Wang} 
\affiliation{Department of Astronomy, Indiana University, 727 East 3rd Street, Bloomington, IN 47405-7105, USA}

 \correspondingauthor{Jace Rusznak}
 \email{jcruszna@iu.edu}

\begin{abstract}

We present a pattern emerging from stellar obliquity measurements in single-star systems: planets with high planet-to-star mass ratios (\mratio$>$ \hotboundarywithourerr) --- such as super-Jupiters, brown dwarf companions, and M-dwarfs hosting Jupiter-like planets --- tend to be aligned, even around hot stars. This alignment represents a \siglevel$\sigma$ deviation from the obliquity distribution observed in systems with lower mass ratios (\mratio$<$ \hotboundarywithourerr), which predominantly include Jupiters and sub-Saturns. The only known outlier system, XO-3, exhibits misalignment confirmed via our newly collected Rossiter-McLaughlin effect measurement ($\lambda=41.8^{+2.1}_{-2.0}$\degrees). However, the relatively large \textit{Gaia} Renormalized Unit Weight Error (RUWE) of XO-3 suggests that it may harbor an undetected binary companion, potentially contributing to its misalignment. Given that tidal realignment mechanisms are weak for hot stars, the observed alignment in high \mratio\, systems is likely \textit{primordial} rather than resulting from tidal interactions. One possible explanation is that only dynamically isolated planets can continue accreting gas and evolve into super-Jupiters while maintaining their primordial alignment. Conversely, planets formed in compact configurations may be unable to grow beyond the gap-opening mass, for which our work suggests an empirical boundary \mratio$=$ \hotboundarywithourerr\, identified between aligned high \mratio\, systems and misaligned low \mratio\, systems, with dynamical instabilities contributing to the diverse spin-orbit misalignments observed in the latter.

\end{abstract}


\section{Introduction} \label{sec:intro}

The angle between the stellar spin axis and the net orbital axis of a planetary system, known as the spin-orbit angle or stellar obliquity, serves as a valuable probe of the dynamical history of planetary systems. The near-coplanar, low-stellar-obliquity configuration of our Solar System suggests a relatively quiescent formation framework \citep{Kant1755, Laplace1796}. However, this aligned configuration is not universal among exoplanetary systems: hot Jupiters exhibit a broad range of stellar spin-orbit angles \citep{Winn2010, Schlaufman2010, Triaud2010, Albrecht2012, WinnFabrycky2015, Albrecht2022, Knudstrup2024}. Specifically, previous work has shown that hot Jupiters orbiting hot, massive stars often display significant spin-orbit misalignment, whereas those around cool, low-mass stars tend to remain aligned \citep{Winn2010, Schlaufman2010}.

\cite{Winn2010} proposed that hot-Jupiter systems may initially be misaligned, with tidal forces realigning cool-star systems with thick convective zones, while hot-star systems with thinner convective zones remain misaligned \citep{Albrecht2012, Xue2014, Valsecchi2014, Li2016, Anderson2021, Wang2021, Rice2022, Spalding2022, zanazzi2024damping,Zanazzi2024a}. \citet{Wu2023HJsNotAlone} suggest that, even if tidal realignment occurs, the initial obliquity distributions --- prior to the potential erasure of misalignments by tidal effects --- may differ between hot Jupiters orbiting cool stars and those around hot stars. High-mass stars with high-mass disks may be more likely to generate multiple Jupiters, which can induce misalignments through planet-planet interactions \citep{Chatterjee2008, Wu2013}. In contrast, low-mass stars with low-mass disks may be more likely to host a single Jupiter, which undergoes quiescent dynamical evolution.

Tentative evidence for alignment in high \mratio\, systems has been noted in previous work. \cite{Hebrard2011} first observed that massive ($\gtrsim$3\mj) planets lack very high obliquities ($>$40\degrees), a finding later emphasized by \citet{Triaud2018}. \cite{Zhou2019HATS70} subsequently reported that the massive planet HATS-70 b has a low stellar obliquity, strengthening this trend. \citet{Gan2024}  highlighted the alignment of TOI-4201 b, an hot Jupiter orbiting an M dwarf, further contributing to this alignment tendency. \citet{Albrecht2022} recently compiled a stellar obliquity catalog including 105 hot-Jupiter systems and found that planets with a mass ratio $\mratio > 2 \times 10^{-3}$ tend to have low stellar obliquities. However, they observed that this alignment trend does not hold for very hot stars ($\teff \geq 7000\, \rm{K}$).
\citet{Hixenbaugh2023} also found suggestive evidence that systems with high \mratio\, tend to be aligned, even around hot stars, with XO-3 being the only outlier (although it is excluded from their Figure 2 due to the y-axis cutoff).

XO-3 b is a massive hot Jupiter ($M_{\text{p}} = 13\,M_{\text{J}}$) with a planet-star mass ratio of $M_{\rm p}/M_{\rm *}=9\times10^{-3}$, orbiting a very hot star ($T_{\text{eff}} = 6,770\,\rm{K}$) on a $3.2$-day orbit. As the first exoplanet measured to have a large spin-orbit misalignment, XO-3 b has been the subject of multiple measurements over time (\citet{Hebrard2008}, $70 \pm 15^\circ$; \citet{Winn2009XO3}, $37.3 \pm 3.7^\circ$; \citet{Hirano2011XO3}, $37.3 \pm 3.0^\circ$), which do not fully agree with each other. Although all previous datasets suffer from systematic noise (see \citet{Hirano2011XO3} and \citet{Worku2022} for detailed discussion), the discrepancies could also reflect true temporal variations in the spin-orbit angle \citep{Rogers2012}.

In this work, with newly collected in-transit spectroscopic observation from the NEID spectrograph, we conducted Rossiter-McLaughlin \citep{Rossiter1924, McLaughlin1924} and Doppler Tomography \citep{Albrecht2007,Collier2010} analyses to re-determine the sky-projected spin-orbit angle of XO-3 and to revisit the population-level alignment trend for systems with high
$M_{\rm p}/M_{\rm *}$. This is the 14$^{\rm th}$ result from the Stellar Obliquities in Long-period Exoplanet Systems (SOLES) survey \citep{Rice2021K2140, WangX2022WASP148, Rice2022WJs_Aligned, Rice2023Q6, Hixenbaugh2023, Dong2023, Wright2023, Rice2023TOI2202, Lubin2023, Hu2024PFS, Radzom2024, Ferreira2024, Wang2024SixWjs}.

Our global modeling results confirm the misaligned configuration for XO-3~b, with $\lambda =\ $\finallambda. This result is consistent with prior high-precision measurements to within 1.3$\sigma$ (\citet{Winn2009XO3}, $37.3 \pm 3.7^\circ$; \citet{Hirano2011XO3}, $37.3 \pm 3.0^\circ$), suggesting no temporal variation in the spin-orbit angle in accordance with the conclusions of \citet{Worku2022}. Furthermore, our analysis of the stellar obliquity distribution as a function of planet-star mass ratio confirms the boundary values previously reported by \citet{Hebrard2011, Triaud2018, Zhou2019HATS70, Albrecht2022, Hixenbaugh2023, Gan2024}: systems with $M_{\rm p}/M_{\rm *}>$\hotboundarywithourerr\ tend to have low obliquities, even around hot stars, with a statistical significance of \siglevel$\sigma$.

The structure of this paper is as follows. In Section~\ref{sec:Obs}, we describe the spectroscopic observations and their reduction. The derivation of atmospheric stellar parameters is outlined in Section~\ref{sec:StPar}. The global modeling is detailed in Section~\ref{sec:Modeling}. A population analysis is provided in Section~\ref{sec:Dis}, with a summary presented in Section~\ref{sec:Summary}.

\section{{In-transit Spectroscopic Observations}} \label{sec:Obs}

In-transit spectroscopic observations of XO-3 were conducted on 2021 October 27 from 4:04--10:27 UT using the NEID spectrograph \citep{Schwab2016} on the WIYN 3.5-meter telescope at Kitt Peak National Observatory in Arizona. We adopted the high-resolution mode (R $\sim$ 110,000) with an exposure time of 720 s. The observations spanned 6.3 hours, encompassing a full transit (3 hours), a 1.8-hour baseline before ingress, and a 1.5-hour baseline after egress. During the observations, atmospheric seeing ranged from 0.6\arcsec{} to 2.0\arcsec{}, with a median seeing of 1.5\arcsec{}, and the median airmass was 1.2.

The NEID spectra were analyzed using the NEID Data Reduction Pipeline (NEID-DRP) v1.3.0\footnote{\url{https://neid.ipac.caltech.edu/docs/NEID-DRP/}}, which utilizes the cross-correlation function (CCF) method to derive radial velocities. The resulting signal-to-noise ratio of the barycentric-corrected radial velocities for re-weighted orders ranged from 149 to 290, with a median value of 239, leading to RV uncertainties ranging from 0.04 to 0.07 km/s.

\begin{figure*}
    \centering
    \makebox[\textwidth][l]{
    \includegraphics[width=1\linewidth]{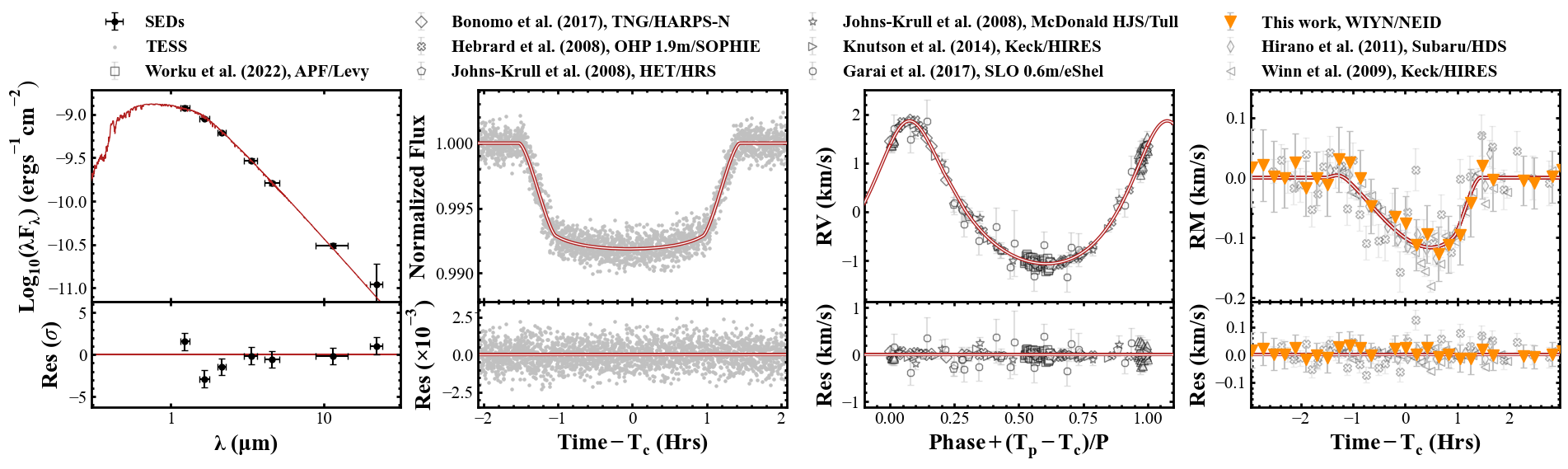}}
    \includegraphics[width=1\linewidth]{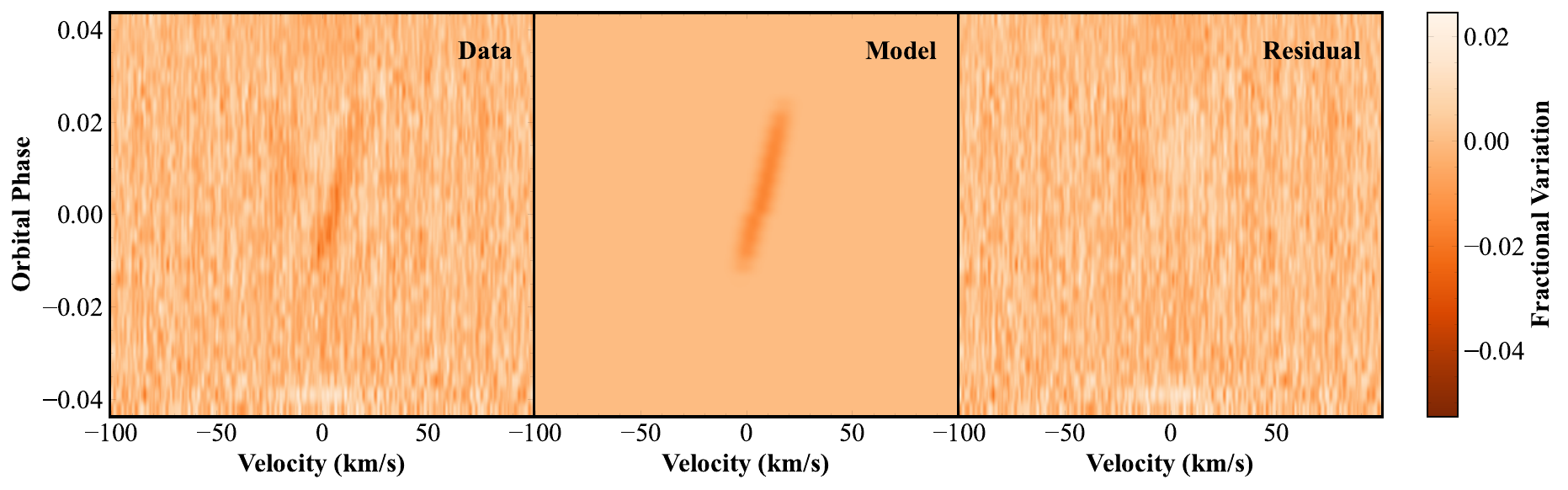}
    \caption{\textit{\textbf{Upper panel}}: Spectral energy distribution, \tess\, transit photometry, out-of-transit RV, and in-transit Rossiter-McLaughlin datasets and associated best-fit EXOFASTv2 models for XO-3 b. NEID radial velocity measurements are represented by inverted orange triangles, and previous measurements \citep{JohnsKrull2008, Hebrard2008, Winn2009XO3, Hirano2011XO3, Knutson2014, Bonomo2017,2017AN....338...35G,Worku2022} are shown in gray. \textit{\textbf{Lower panel}}: Doppler Tomography signal for XO-3, obtained during transit. Our spectra captured by NEID produce the Doppler tomography seen in the left panel, while the best-fit model and associated residuals are shown in the middle and right panels, respectively.}
    \label{Fig:modelling}
\end{figure*}

\begin{deluxetable}{cccc}
\tablecaption{NEID Radial Velocities for the XO-3 System. \label{tab:NEID}}        
\tablewidth{0pt}                                                     
\tablehead{
\colhead{Time (BJD$_{\rm TDB}$)} & \colhead{RV (km s$^{-1}$)} & \colhead{Error (km s$^{-1}$)}
}                         
\startdata                                                            
2459514.68763608 & -11.1265 & 0.0747 &  \\                           
2459514.69556828 & -11.1616 & 0.0647 &  \\                           
2459514.70442987 & -11.1867 & 0.0599 &  \\                           
2459514.71321440 & -11.2340 & 0.0591 &  \\                           
2459514.72184409 & -11.2649 & 0.0490 &  \\                           
2459514.73043128 & -11.2660 & 0.0471 &  \\                           
2459514.73908392 & -11.3376 & 0.0456 &  \\                           
2459514.74766736 & -11.3481 & 0.0447 &  \\                           
2459514.75625436 & -11.3862 & 0.0514 &  \\                           
2459514.76509763 & -11.3719 & 0.0544 &  \\                           
2459514.77376901 & -11.4033 & 0.0514 &  \\                           
2459514.78217161 & -11.4558 & 0.0465 &  \\                           
2459514.79105276 & -11.5296 & 0.0452 &  \\                           
2459514.80988314 & -11.6043 & 0.0484 &  \\                           
2459514.81882804 & -11.6420 & 0.0460 &  \\                           
2459514.82741268 & -11.7027 & 0.0426 &  \\                           
2459514.83577554 & -11.7092 & 0.0481 &  \\                           
2459514.84486433 & -11.7681 & 0.0454 &  \\                           
2459514.85303856 & -11.7761 & 0.0474 &  \\                           
2459514.86193159 & -11.7853 & 0.0493 &  \\                           
2459514.87061424 & -11.7555 & 0.0566 &  \\                           
2459514.87959117 & -11.7190 & 0.0540 &  \\                           
2459514.88814168 & -11.7648 & 0.0529 &  \\                           
2459514.91309489 & -11.8330 & 0.0410 &  \\                           
2459514.92172524 & -11.8595 & 0.0428 &  \\                           
2459514.93557633 & -11.8848 & 0.0431 &  \\                           
2459514.94383404 & -11.8944 & 0.0409 &  \\                           
\enddata
\tablenotetext{}{(This table is in machine-readable form in the online manuscript.)}
\end{deluxetable}

\subsection{\tess\, Photometry}

In this work, we adopted the Transiting Exoplanet Survey Satellite \citep[\tess;][]{Ricker2015} Presearch Data Conditioning simple aperture photometry \citep[PDCSAP;][]{Smith2012, Stumpe2012, Stumpe2014} data, which was processed by the \tess\ Science Processing Operations Center \citep[SPOC;][]{Jenkins2016} at NASA Ames Research Center. The data was downloaded using the \texttt{lightkurve} \citep{Lightkurve2018} Python package and includes 12 transits observed with 2-minute cadence across Sectors 19, 59, and 73, covering a four year time frame.

\section{{Atmospheric Stellar Parameters Derivation}} \label{sec:StPar}

To derive stellar atmospheric parameters such as \teff, \feh, \logg, and \vsini\, for XO-3, we performed synthetic spectral fitting using \texttt{iSpec}\footnote{\url{https://github.com/marblestation/iSpec}} \citep{Blanco2014, Blanco2019} on the co-added out-of-transit NEID spectra (SNR=258). We adopted the SPECTRUM radiative transfer code \citep{Gray1994}, the MARCS atmosphere model \citep{gustafsson2008_MARCS}, and the sixth version of the GES atomic line list \citep{Heiter2021_GES}, all embedded in \texttt{iSpec}, to generate synthetic spectral models. Specific lines, including the wings of the H$\alpha$, H$\beta$, and Mg I triplet lines, as well as Fe I and Fe II lines, were selected to expedite the fitting process. We fixed the linear limb darkening coefficient and resolution at 0.6 and 110,000, respectively. The microturbulent velocity was treated as a free parameter, while the macroturbulent velocity was calculated using the empirical relation reported by \citet{Doyle2014Vmac}. Using \texttt{iSpec}, we applied a nonlinear least-squares Levenberg–Marquardt algorithm \citep{more2006levenberg} to reduce the differences between the NEID and synthetic spectra. Upon achieving convergence, the uncertainty for each parameter was derived from the covariance matrix. The resulting parameters are listed in Table~\ref{tab:combined}.

\begin{figure*}
    \centering
    \includegraphics[width=1\linewidth, trim= 1 1 1 2, clip]{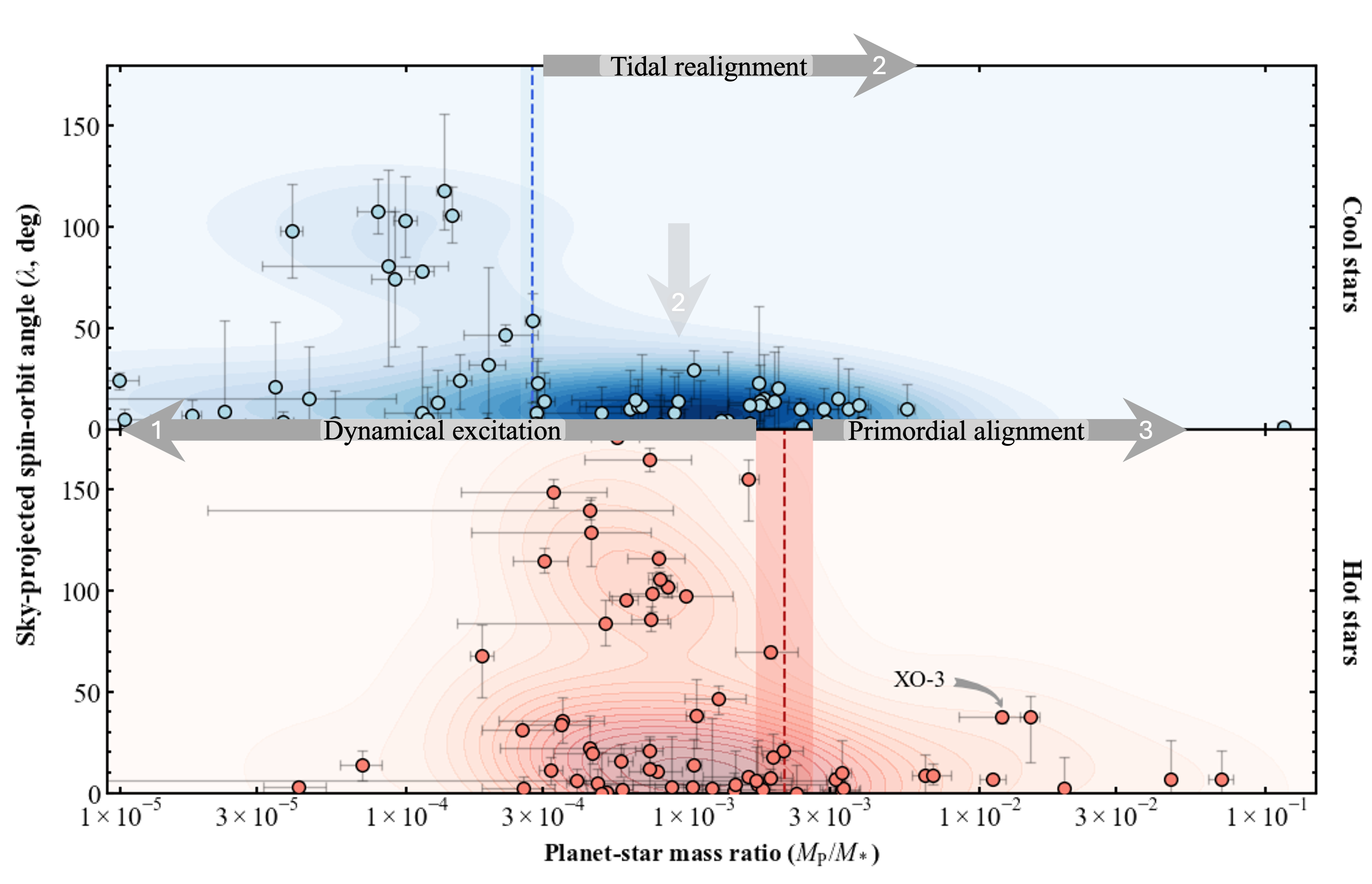}
    \caption{
    \textbf{Upper panel}: The sky-projected spin-orbit angle ($|\lambda|$) distribution for cool-star systems as a function of plant-to-star mass ratio. The contours represent the kernel density estimation of stellar obliquity measurements in the $\mratio–|\lambda|$ space. The vertical blue dashed line marks the most likely boundary between the misalignment of low \mratio\, systems and the alignment of high \mratio\, systems. \textbf{Lower panel}: Similar to the upper panel, but for hot-star systems. The location of XO-3 is indicated by a curved gray arrow.
    The left-oriented arrow, marked as `1', indicates that dynamical excitation can lead to misalignment in both cool-star and hot-star systems with \mratio $<$ \hotboundarywithourerr. However, for cool stars with thick convective envelopes, misaligned systems with \coolboundarywithourerr $<$ \mratio $<$ \hotboundarywithourerr are tidally damped, as shown by arrows marked as `2'. In systems with high mass ratios, represented by the `3' arrow, primordial alignment dominates the stellar obliquity distribution, with no dynamical excitation causing misalignment. The data used to create this plot can be found via \href{https://github.com/wangxianyu7/Data_and_code/tree/main/HighMassRatioSystemsTendToHaveLowStellarObliuiqties}{this link}.}

    \label{Fig:spinorbit}

\end{figure*}

\section{Global Modeling} \label{sec:Modeling}

To maintain consistency across both stellar and planetary parameters, we adopted EXOFASTv2\footnote{Main branch: \url{https://github.com/jdeast/EXOFASTv2}; \\
Version used in this work: \url{https://github.com/wangxianyu7/EXOFASTv2}.} \citep{Eastman2017, Eastman2019} to perform a global fit to the XO-3 spectral energy distribution (SED), transits, transit mid-times, and radial velocities (including the in-transit Rossiter-McLaughlin/Doppler Tomography measurements).

\begin{itemize}
\item MIST+SED modeling

To derive the system's stellar parameters -- particularly \mstar, \rstar, \logg, \feh, and stellar age -- we conducted SED fitting with the MESA Isochrones \& Stellar Tracks \citep[MIST;][]{Choi2016mist, Dotter2016mist} framework. The photometric magnitudes of XO-3 were compiled from multiple catalogues, including 2MASS \citep{Cutri2003}, WISE \citep{Cutri2014AllWISE}, \textit{TESS} \citep{Ricker2015}, and \textit{Gaia} DR3 \citep{GaiaCollaboration2023}. Gaussian priors on \teff\, and \feh\, were based on values from the NEID spectral analysis. We adopted an upper limit on the V-band extinction from \cite{Schlafly2011} and a Gaussian prior on parallax from \textit{Gaia} DR3 \citep{GaiaCollaboration2023}. Systematic uncertainty floors of 2.4\% for bolometric flux and 2\% for effective temperature, as suggested by \cite{Tayar2022}, were applied.

\item Photometry+out-of-transit RV modeling

The planetary parameters were derived from photometry and out-of-transit RV datasets. To expedite the modeling process, the \tess\, photometry and ground-based transit mid-times from the literature, as reported by \cite{Ivshina2022}, were fitted simultaneously. \tess\, photometry provide a high-precision transit profile, and literature transit mid-times help to maintain an accurate ephemeris. The \tess\ light curve was detrended using the cosine method implemented in \texttt{wotan}\footnote{\url{https://github.com/hippke/wotan}} \citep{Hippke2019}.
Literature radial velocity data, which we leveraged to accurately constrain the planetary mass, eccentricity, and longitude of periastron, were obtained from the following spectrographs: 
the High Resolution Spectrograph (HRS) on the Hobby-Eberly Telescope and the Tull Coudé spectrograph on the Harlan J. Smith telescope \citep{Tull1995, Tull1998}; the High Dispersion Spectrograph (HDS) on the Subaru telescope \citep{Noguchi2002}; the Spectrographe pour l'Observation des Ph$\mathrm{\acute{e}}$nom$\mathrm{\grave{e}}$nes des Int$\mathrm{\acute{e}}$rieurs stellaires et des Exoplan$\mathrm{\grave{e}}$tes (SOPHIE) on the 1.93m reflector telescope at the Haute-Provence Observatory \citep{Perruchot2008}; the northern High Accuracy Radial velocity Planet Searcher (HARPS-N) on the Telescopio Nazionale Galileo \citep{Cosentino2012, Cosentino2014}; the Levy spectrograph on the Automated Planet Finder (APF) at Lick Observatory \citep{Vogt2014}; the HIgh Resolution Echelle Spectrometer (HIRES) on Keck I \citep{Vogt2014}; and the eShel spectrograph at Stara Lesna Observatory \citep{Eversberg2016}.
Uniform priors on planetary period, time of conjunction, planetary radius in stellar radii ($R_P/R_*$), semi-major axis in stellar radii ($a/R_*$), transformed eccentricity and longitude of periastron (\(\secosw\) and \(\sesinw\)), and impact parameter were applied. The priors on the quadratic limb-darkening coefficients (LDC) for \tess\, transits were interpolated using the \cite{Claret2011} model based on the \teff, \feh, and \logg.

\item RM modeling

To accurately model the RMs from \cite{Hebrard2008}, \cite{Winn2010}, \cite{Hirano2011XO3}, and this work, we implemented one of the most realistic RM models presented by \cite{Hirano2011}, which accounts for stellar rotation, macroturbulence, thermal broadening, pressure broadening, and instrumental broadening. The $V$-band quadratic limb darkening coefficients were adopted, with priors interpolated using the \cite{Claret2011} model based on \teff, \feh, and \logg. A Gaussian prior on \vsini\, was adopted based on the value derived from NEID spectral analysis (see Section~\ref{sec:StPar}). The prior on the macroturbulence velocity (\vzeta) was applied using the empirical equation described in \cite{Doyle2014Vmac}. Uniform priors on the Gaussian dispersion (\vbeta) and Lorentzian dispersion (\vgamma) were set to 2.5--4.5 \kms and 0.5--1.5 \kms, respectively, consistent with the typical ranges presented by \cite{Hirano2011}.

\item Doppler tomography modeling

Because of the fast rotation of XO-3 (\vsini$\sim$18 $\text{km s}^{-1}$), the line profile variation caused by the transiting planet is significant. This enables a derivation of the projected spin-orbit angle using the Doppler tomography technique \citep{Albrecht2007,Collier2010,Zhou2016,Johnson2017}. To extract the Doppler tomography signal from NEID Level 2 data, we processed the CCF files by shifting their velocity coordinates to the stellar rest frame via subtraction of the systemic velocity. We then created a reference stellar line profile by averaging the out-of-transit CCFs. Subtracting this reference from the CCF of each observation isolated the DT signal of the planetary transit. Instead of adopting \vbeta, \vgamma, and \vzeta, we introduced \vvline\, to represent the average line width excluding the rotational broadening effect. A uniform prior of 2--10 \kms\, was adopted for $v_{\rm line}$. The prior on $v\sin i_\star$ remains the same as in the RM modeling.

\item Optimization and Uncertainty Sampling

We used EXOFASTv2 to simultaneously perform the aforementioned modeling and derive robust parameter uncertainties. First, EXOFASTv2 searched for initial guesses for floating parameters using the AMOEBA algorithm \citep{NelderMead1965}. Once AMOEBA found local minima, the Parallel Tempering Differential Evolution Markov Chain Monte Carlo \citep[DE-MCMC;][]{ter2006markov} process was initiated with the best guesses. The MCMC procedure was considered converged when the Gelman-Rubin diagnostic \citep[$\hat{R}$;][]{Gelman1992} fell below 1.01 and the count of independent draws surpassed 1000. The resulting parameters are listed in Table~\ref{tab:combined}.

\item Global modeling with different data combinations

To examine how sensitive the derived parameters are to different dataset combinations, we conducted four distinct global fits using varying datasets. All fits adopted \tess\, transits, literature transit mid-times, and RVs, with the differences arising from the usage of RMs and DT data. In the first case, we performed a fit using all literature RMs combined with the DT data from this work (denoted as ``Literature RMs + NEID DT''). In the second case, we included all RMs, including those newly collected in this work (``Literature RMs + NEID RM''). In the third case, we used only our NEID DT data without any additional RMs (``NEID DT''). In the last case, we used only the NEID RM data (``NEID RM''). To maintain consistency with the results from RM modeling in the literature, we adopt the results from the “Literature RMs + NEID RM” fit, which yields an RM SNR of 338 \citep{Kipping2023,Kipping2024} and $\lambda=$\finallambda. The results of each of the four fits are listed in Table~\ref{tab:combined}. 

We found that the resulting system parameters are in good agreement with values derived in previous studies (differences within $\lesssim2\sigma$), except for the stellar effective temperature and radius. The stellar \teff\, derived from our analysis is consistent (within 1$\sigma$) with the values reported by \cite{2014A&A...570A..80T}, \textit{Gaia} DR2 \citep{GaiaCollaboration2018}, \textit{Gaia} DR3 \citep{GaiaCollaboration2023}, and the TIC catalog \citep{Stassun2018}, but differs by 2.4$\sigma$ from the value reported by \cite{JohnsKrull2008}. This discrepancy may be attributable to differences in data quality and spectral analysis software. For stellar radius, the 2.8$\sigma$ inconsistency may originate from differences in parallax between \textit{Gaia} DR2 and DR3. Previous studies (e.g. \citet{Worku2022}) adopted the \textit{Gaia} DR2 \citep{GaiaCollaboration2016} parallax (4.66$\pm$0.06 mas), which differs significantly from the \textit{Gaia} DR3 parallax (4.86$\pm$0.03 mas) used in this work. Notably, our $\lambda$ and \vsini\, agree with values from previous accurate measurements (\citealt{Winn2009XO3}: $\lambda$=37.3$\pm$3.7\degrees, \vsini=17.0$\pm$1.2\kms; \citealt{Hirano2011XO3}: $\lambda$=37.3$\pm$3.0\degrees, \vsini=18.54$\pm$0.17\kms) within 1.3$\sigma$.

In some cases, the true stellar obliquity $\psi$ can be determined if the stellar rotational velocity $v$ is inferred from photometric variability caused by long-lived starspots on the surface of the host star (e.g. \citet{Mancini2015}, \citet{Bourrier2018}, \citet{Stefansson2022}). Therefore, we applied the autocorrelation function embedded in \texttt{SpinSpotter} \citep{Holcomb2022} to the \tess\, light curve. We did not detect any significant periodic signal; therefore, the true stellar obliquity of XO-3 cannot yet be determined.
\end{itemize}

\section{{Population Analysis}} \label{sec:Dis}

Previous studies (e.g.  \citet{ Hebrard2011}, \citet{Triaud2018}, \citet{Zhou2019HATS70}, \citet{Albrecht2022}, \citet{Hixenbaugh2023}, \citet{Gan2024}) reported tentative evidence for a trend toward alignment in massive planet (\mplanet$\gtrsim$ 3\mj) or high \mratio\, (\mratio$>$\hotboundarywithourerr) systems . However, historically, a limited sample size has made the statistical validation of this trend uncertain. Recently, the sample of high \mratio\, systems has been rapidly expanding, benefiting from \tess, which has uncovered numerous such systems and facilitated a series of follow-up stellar obliquity observations (e.g.,  LP 261-75A b, \citet{Brady2024}; TOI-2119 b, \citet{Doyle2024}; TOI-2145, \citet{Dong2024}; TOI-2533, \citet{Schmidt2023, Ferreira2024}; TOI-3362, \citet{Dong2021, Espinoza2023}; TOI-4603, \citet{Khandelwal2023}; TIC 241249530, \citet{Gupta2024}; GPX-1, \citet{Benni2021, Giacalone2024}). We revisit the \mratio trend in light of these new measurements.

\begin{itemize}

\item Sample construction 

In this work, using literature stellar obliquities from TEPcat\footnote{Accessed on November 17, 2024, at\\ \url{https://www.astro.keele.ac.uk/jkt/tepcat/obliquity.html}.} \citep{Southworth2011TEPcat} and recent studies including \cite{Knudstrup2024}, \cite{Dong2024}, \cite{Brady2024}, and \cite{Doyle2024}, we conducted a population-level analysis to re-investigate the tentative alignment trend in single-star, high \mratio\, systems. For each system, we used the preferred values for $\lambda$ and \teff\ from the TEPCat catalog, while \mstar\, and \mplanet\, were sourced from the Planetary Systems Composite Data table in the NASA Exoplanet Archive\footnote{\url{https://exoplanetarchive.ipac.caltech.edu/cgi-bin/TblView/nph-tblView?app=ExoTbls&config=PSCompPars}}. To maintain the purity of the sample, following the first three criteria of the sample construction procedure outlined in \href{https://iopscience.iop.org/article/10.3847/2041-8213/ad7469#apjlad7469app2}{Appendix B} of \cite{Wang2024SixWjs}, we included only stellar obliquity measurements obtained using the Rossiter-McLaughlin or Doppler tomography techniques, while excluding low-quality and contested measurements, as well as systems containing binary or multiple stars. We also excluded systems with planetary masses that have only upper or lower limits.

This procedure resulted in \nall\, systems, including \nbd\, brown dwarf systems (CoRoT-3, \citet{Triaud2009}; GPX-1, \citet{Giacalone2024}; KELT-1, \citet{Siverd2012}; TOI-2119, \citet{Doyle2024}; TOI-2533, \citet{Ferreira2024}; WASP-30, \citet{Triaud2013WASP30}, and XO-3). Among them, \ncool\, systems include cool host stars (\teff$<$ 6100 K), while \nhot\, systems include hot host stars (\teff$\geq$ 6100 K). The resulting sky-projected stellar obliquity distribution is shown in Figure~\ref{Fig:spinorbit}. There are clear differences in the obliquity distributions between systems with low and high \mratio\, ratios in both the hot and cool star samples.

\item Empirical boundary

To determine the empirical threshold that distinguishes the $\lambda$ distributions of systems with low and high $M_{\rm p}/M_{\rm *}$ ratios, we applied a Anderson-Darling test \citep[AD;][]{Scholz1987KSampleAT} implemented in \texttt{scipy} \citep{virtanen2020scipy}. The null hypothesis assumes that the stellar obliquity distributions on both sides of the boundary are identical. For each iteration, random $\lambda$ values were drawn for the systems, incorporating the reported Gaussian uncertainties for each measurement. The boundary value was varied logarithmically from $10^{-4}$ to $10^{-1}$ with a step size of 0.001, and the boundary with most significant AD statistic was recorded. This process was repeated 1,000 times to generate the AD statistic distribution as a function of \mratio, which was fitted with a Gaussian profile to determine the median value and 1$\sigma$ uncertainty for boundaries. This procedure was applied separately to the cool-star and hot-star systems, resulting in \mratio\, boundaries of \coolboundary\, and \hotboundary\,, respectively. 

\item Statistical significance

For the cool star sample (as shown in the top panel of Figure~\ref{Fig:spinorbit}), there are \ncoollowmassratio\ systems with $M_{\rm p}/M_{*}<$\coolboundarywithourerr\, spanning a range of spin-orbit orientations. In contrast, \ncoolhighmassratio\ systems with $M_{\rm p}/M_{*}\geq$\coolboundarywithourerr\, are generally well aligned. Considering the similar sample sizes of the low \mratio\, and high \mratio\, groups, we applied the AD test and a Monte Carlo approach to compare the distributions of the two groups while accounting for stellar obliquity uncertainties. The null hypothesis is that the two samples originate from the same distribution. In each iteration, Gaussian noise was introduced to simulate measurement errors, and the p-value of the AD test was calculated. This process was repeated 100,000 times to build a robust distribution of AD p-values. The results show that 97.8\% of the resultant p-values are less than 0.05, strongly rejecting the null hypothesis and indicating that, for the stellar obliquity distribution of cool stars, low \mratio\, and high \mratio\, systems are distinct.

For the hot star sample (as shown in the bottom panel of Figure~\ref{Fig:spinorbit}), there are \nhotlowmassratio\, systems with \mratio$<$\hotboundarywithourerr\,, spanning a range of spin-orbit orientations, while \nhothighmassratio\, systems with \mratio$\geq$\hotboundarywithourerr\, tend to have low stellar obliquity. Similar to the cool star sample, a comparable AD test analysis was conducted for the hot star sample, showing that 85.9\% of the resultant p-values are less than 0.05, suggesting an intrinsic difference between low \mratio\ and high \mratio\ systems with hot stars.
Due the highly uneven sample size, we also employed a bootstrapping method to assess the significance of the observed alignment trend in high \mratio\, systems. Following \citet{WangX2022WASP148}, \citet{Rice2022WJs_Aligned}, and \cite{Wang2024SixWjs}, we define ``misaligned" systems as those with $|\lambda| > 10^{\circ}$ and $\lambda$ deviating from $0^\circ$ at a $3\sigma$ significance level. Specifically, we iteratively draw random samples of \nhothighmassratio\,  $|\lambda|$ values from the hot-star low \mratio\ sample without replacement to determine how many are misaligned. This process is repeated 100,000 times to establish a stable distribution for the number of misaligned systems. The results demonstrate that the difference in the spin-orbit distribution between hot-star low \mratio\ and high \mratio\ systems is significant at the \siglevel$\sigma$ level.

The boundaries derived from our work are similar to those reported by \cite{Albrecht2022}. However, \cite{Albrecht2022} found that the suggestive alignment in high \mratio\, systems does not hold for very hot stars (\teff$ > 7000$ K). We, therefore, examined the misaligned systems with very hot stars (\teff $>$ 7000 K) shown in the lower panel of Figure 9 in \cite{Albrecht2022}, and found that all misaligned systems with \mratio$\geq$\hotboundarywithourerr\, are either in binary systems (Kepler-13, \citet{2011ApJ...736...19B}; KELT-19, \citet{2018AJ....155...35S}) or have only upper limits on their planetary masses (HAT-P-70, \citet{2019AJ....158..141Z}; WASP-167, \citet{2017MNRAS.471.2743T}).

\item XO-3: Exception

Among single-star systems with high \mratio --- which are typically well-aligned --- XO-3 b stands out as a notable exception. It possesses an unusually large planet-star mass ratio (\mratio=$9\times10^{-3}$) and exhibits a high stellar obliquity ($\lambda =\ $\finallambda), presenting a significant challenge in understanding its dynamical history \citep{JohnsKrull2008, Hebrard2008, Winn2009XO3, Hirano2011XO3, Worku2022}. We, therefore, consider the possibility that the system harbors an undetected stellar companion that may have misaligned the system \citep[e.g.][]{Wu2003, Fabrycky2007, Hjorth2021, Espinoza2023, Suyubo2024}.

To search for a potential stellar companion, we first applied the RV planet search pipeline, \texttt{rvsearch}\footnote{\url{https://github.com/California-Planet-Search/rvsearch}} \citep{Rosenthal2021} based on RadVel \citep{Fulton2018RadVel}, to the residual RV data obtained after removing the signal induced by XO-3 b. Although \citet{Hirano2011XO3} identified a tentative linear trend in radial velocity data from Subaru/HDS and OHP/SOPHIE, our analysis found no significant signals with a false alarm probability below 0.1. This suggests either the absence of a nearby companion or an insufficient number or precision of RVs to detect one. Unfortunately, due to the lack of high-precision astrometry measurements in the Hipparcos catalog \citep{HIPPARCOS1997}, Hipparcos+\textit{Gaia} astrometry detection cannot be achieved.

Previous imaging observations of XO-3 did not find a physically associated stellar companion (Lucky Imaging by AstraLux Norte \citep{Bergfors2013, Wollert2015}; direct imaging by Keck/NIRC2 \citep{Ngo2015}). However, \textit{Gaia} DR3 \citep{GaiaCollaboration2023} reports that XO-3 has a Renormalized Unit Weight Error RUWE; \citep[RUWE;][]{Lindegren2018, Lindegren2021} of 1.25, suggesting a poor astrometric solution, whereas RUWE is typically close to unity for single stars with reliable astrometric data. A high RUWE ($>1.2$) indicates a strong likelihood of a nearby unresolved stellar companion \citep{Belokurov2020, Krolikowski2021}, although other factors, such as circumstellar material \citep{Fitton2022RNAAS}, can also contribute to an elevated RUWE. The potential presence of a stellar companion can be tested with the upcoming \textit{Gaia} DR4 epoch data. If the binary nature of XO-3 is confirmed, it would no longer represent the only known misaligned single-star system with \mratio$\geq$\hotboundarywithourerr. The statistical significance of alignment in high \mratio\, single-star systems would, in this case, increase from \siglevel$\sigma$\, to 4.1$\sigma$.

\end{itemize}


\providecommand{\bjdtdb}{\ensuremath{\rm {BJD_{TDB}}}}
\providecommand{\tjdtdb}{\ensuremath{\rm {TJD_{TDB}}}}
\providecommand{\feh}{\ensuremath{\left[{\rm Fe}/{\rm H}\right]}}
\providecommand{\teff}{\ensuremath{T_{\rm eff}}}
\providecommand{\teq}{\ensuremath{T_{\rm eq}}}
\providecommand{\ecosw}{\ensuremath{e\cos{\omega_*}}}
\providecommand{\esinw}{\ensuremath{e\sin{\omega_*}}}
\providecommand{\msun}{\ensuremath{\,M_\Sun}}
\providecommand{\rsun}{\ensuremath{\,R_\Sun}}
\providecommand{\lsun}{\ensuremath{\,L_\Sun}}
\providecommand{\mj}{\ensuremath{\,M_{\rm J}}}
\providecommand{\rj}{\ensuremath{\,R_{\rm J}}}
\providecommand{\me}{\ensuremath{\,M_{\rm E}}}
\providecommand{\re}{\ensuremath{\,R_{\rm E}}}
\providecommand{\fave}{\langle F \rangle}
\providecommand{\fluxcgs}{10$^9$ erg s$^{-1}$ cm$^{-2}$}
\startlongtable
\begin{deluxetable*}{lcccccccccc}
\tabletypesize{\scriptsize}
\tablewidth{10pt}
\tablecaption{Median values and 68\% confidence interval for combined parameters.}
\tablehead{\colhead{~~~Parameter} & \colhead{Description} & \colhead{Values} & \colhead{} & \colhead{} & \colhead{}}
\startdata
\multicolumn{5}{l}{\textbf{Stellar parameters from iSpec fit:}}\\ 
$T_{\rm eff}$ \dotfill & Effective Temperature (K) \dotfill & ${6784\pm141}$ &&  &  \\
$[{\rm Fe/H}]$ \dotfill & Metallicity (dex) \dotfill & $-0.23\pm0.09$ &  &&  \\
$\log{g}$ \dotfill & Surface Gravity (cgs) \dotfill & $4.05\pm0.32$ && & \\
$\vsini$ \dotfill & Projected rotational velocity (m/s) \dotfill & $19.28\pm1.61$ & &  &  \\
&&&&&&&\\
\hline\hline
    && Literature RMs & Literature RMs & NEID DT & NEID RM\\
    && +        & +  &   &  \\
    && NEID DT  &  NEID RM &   &  \\
                &&   & (adopted)  &   &  \\
\hline
\multicolumn{5}{l}{\textbf{EXOFASTv2 fit:}}\\ 
\multicolumn{5}{l}{\textbf{\scriptsize{Stellar Parameters:}}}\\ 
$M_*$ \dotfill & Mass (\msun) \dotfill & ${1.41}^{+0.067}_{-0.069}$ & ${1.409}^{+0.066}_{-0.07}$ & ${1.433}^{+0.066}_{-0.07}$ & ${1.424}^{+0.066}_{-0.07}$ \\
$R_*$ \dotfill & Radius (\rsun) \dotfill & ${1.52}^{+0.029}_{-0.03}$ & ${1.508}^{+0.029}_{-0.03}$ & ${1.524}^{+0.029}_{-0.03}$ & ${1.515}^{+0.029}_{-0.03}$ \\
$L_*$ \dotfill & Luminosity (\lsun) \dotfill  & ${1.582}^{+0.034}_{-0.032}$ & ${1.583}^{+0.034}_{-0.033}$ & ${1.57}^{+0.034}_{-0.032}$ & ${1.575}^{+0.034}_{-0.033}$ \\
$F_{Bol}$ \dotfill & Bolometric Flux (cgs) \dotfill & ${3.17}^{+0.25}_{-0.23}$ & ${3.09}^{+0.24}_{-0.23}$ & ${3.47}^{+0.32}_{-0.3}$ & ${3.3}^{+0.28}_{-0.27}$ \\
$\rho_*$ \dotfill & Density (cgs) \dotfill & ${0.565}^{+0.024}_{-0.022}$ & ${0.578}^{+0.024}_{-0.023}$ & ${0.569}^{+0.024}_{-0.023}$ & ${0.577}^{+0.024}_{-0.023}$ \\
$\log{g}$ \dotfill & Surface Gravity (cgs) \dotfill & ${4.223}^{+0.014}_{-0.015}$ & ${4.23}^{+0.014}_{-0.014}$ & ${4.227}^{+0.015}_{-0.015}$ & ${4.23}^{+0.014}_{-0.015}$ \\
$T_{\rm eff}$ \dotfill & Effective Temperature (K) \dotfill & ${6790}^{+100}_{-110}$ & ${6770}^{+100}_{-100}$ & ${6940}^{+130}_{-130}$ & ${6870}^{+120}_{-120}$ \\
$[{\rm Fe/H}]$ \dotfill & Metallicity (dex) \dotfill & ${-0.131}^{+0.089}_{-0.11}$ & ${-0.122}^{+0.088}_{-0.11}$ & ${-0.166}^{+0.097}_{-0.12}$ & ${-0.146}^{+0.092}_{-0.12}$ \\
Age \dotfill & Age (Gyr) \dotfill & ${1.25}^{+0.54}_{-0.4}$ & ${1.2}^{+0.54}_{-0.39}$ & ${1.09}^{+0.47}_{-0.35}$ & ${1.11}^{+0.49}_{-0.37}$ \\
$EEP$ \dotfill & Equal Evolutionary Phase \dotfill & ${339.7}^{+8.4}_{-10.0}$ & ${338.2}^{+8.9}_{-10.0}$ & ${337.2}^{+8.6}_{-11.0}$ & ${337.0}^{+8.8}_{-11.0}$ \\
$A_V$ \dotfill & V-band Extinction (mag) \dotfill & ${0.16}^{+0.15}_{-0.11}$ & ${0.14}^{+0.15}_{-0.1}$ & ${0.16}^{+0.15}_{-0.11}$ & ${0.18}^{+0.15}_{-0.12}$ \\
$\varpi$ \dotfill & Parallax (mas) \dotfill & ${4.74}^{+0.064}_{-0.064}$ & ${4.745}^{+0.063}_{-0.063}$ & ${4.731}^{+0.064}_{-0.064}$ & ${4.738}^{+0.064}_{-0.064}$ \\
$d$ \dotfill & Distance (pc) \dotfill & ${211.0}^{+2.9}_{-2.8}$ & ${210.8}^{+2.8}_{-2.8}$ & ${211.4}^{+2.9}_{-2.8}$ & ${211.1}^{+2.9}_{-2.8}$ \\
\multicolumn{5}{l}{\textbf{\scriptsize{Rossiter-McLaughlin and Doppler Tomography Parameters:}}}\\ 
$\lambda$ \dotfill &   Projected spin-orbit angle (deg) \dotfill & ${39.5}^{+1.4}_{-1.4}$ & ${41.8}^{+2.1}_{-2.0}$ &${39.5}^{+1.6}_{-1.6}$ & ${37.1}^{+5.1}_{-4.8}$ \\
$\vsini$ \dotfill & Projected rotational velocity (m/s) \dotfill & ${18050}^{+230}_{-230}$ & ${18630}^{+240}_{-240}$ & ${17290}^{+380}_{-380}$ & ${18550}^{+490}_{-500}$ \\
$\vvline$ \dotfill &  Unbroadened line width (m/s)\dotfill & ${8590}^{+540}_{-510}$ & $\mathrm{...}$ & ${8250}^{+550}_{-520}$ & $\mathrm{...}$ \\
$\vbeta$ \dotfill & Gaussian dispersion (m/s) \dotfill & ${3790}^{+520}_{-750}$ & ${3640}^{+600}_{-710}$ & ${3500}^{+680}_{-680}$ & ${3490}^{+690}_{-680}$ \\
$\vgamma$ \dotfill & Lorentzian dispersion (m/s) \dotfill & ${1060}^{+300}_{-360}$ & ${1010}^{+340}_{-340}$ & ${1000}^{+340}_{-340}$ & ${1000}^{+340}_{-340}$ \\
$\vzeta$ \dotfill & Macroturbulence dispersion (m/s) \dotfill &  ${6170}^{+240}_{-490}$ & ${6150}^{+260}_{-510}$ & ${4300}^{+1500}_{-1500}$ & ${5640}^{+630}_{-1100}$ \\
\multicolumn{5}{l}{\textbf{\scriptsize{Planetary Parameters:}}}\\ 
$P$ \dotfill & Period (days) \dotfill & ${3.19152313}^{+1.5e-07}_{-1.5e-07}$ & ${3.19152308}^{+1.5e-07}_{-1.4e-07}$ & ${3.19152307}^{+1.5e-07}_{-1.5e-07}$ & ${3.19152306}^{+1.5e-07}_{-1.5e-07}$ \\
$R_P$ \dotfill & Radius (\rj) \dotfill & ${1.317}^{+0.027}_{-0.027}$ & ${1.307}^{+0.027}_{-0.028}$ & ${1.32}^{+0.027}_{-0.028}$ & ${1.312}^{+0.027}_{-0.028}$ \\
$M_P$ \dotfill & Mass (\mj) \dotfill & ${13.03}^{+0.41}_{-0.43}$ & ${13.03}^{+0.41}_{-0.43}$ & ${13.18}^{+0.4}_{-0.44}$ & ${13.13}^{+0.4}_{-0.43}$ \\
$T_C$ \dotfill & Time of conjunction - 2457417\dotfill & ${0.98762}^{+0.00011}_{-0.00011}$ & ${0.9876}^{+0.00011}_{-0.00011}$ & ${0.98762}^{+0.00011}_{-0.00011}$ & ${0.98761}^{+0.00011}_{-0.00011}$ \\
$a$ \dotfill & Semi-major axis (AU) \dotfill & ${0.04771}^{+0.00074}_{-0.00079}$ & ${0.0477}^{+0.00073}_{-0.0008}$ & ${0.04796}^{+0.00072}_{-0.0008}$ & ${0.04787}^{+0.00072}_{-0.00079}$ \\
$i$ \dotfill & Inclination (Degrees) \dotfill & ${83.57}^{+0.18}_{-0.17}$ & ${83.67}^{+0.17}_{-0.17}$ & ${83.6}^{+0.18}_{-0.18}$ & ${83.66}^{+0.18}_{-0.18}$ \\
$e$ \dotfill & Eccentricity \dotfill & ${0.2795}^{+0.0026}_{-0.0024}$ & ${0.279}^{+0.0024}_{-0.0023}$ & ${0.2785}^{+0.0026}_{-0.0024}$ & ${0.2784}^{+0.0025}_{-0.0023}$ \\
$\omega_*$ \dotfill & Arg of periastron (Degrees) \dotfill & ${-11.08}^{+0.86}_{-0.95}$ & ${-10.77}^{+0.76}_{-0.82}$ & ${-10.54}^{+0.88}_{-0.95}$ & ${-10.51}^{+0.83}_{-0.89}$ \\
$R_P/R_*$ \dotfill & Radius of planet in stellar radii \dotfill & ${0.08906}^{+0.00026}_{-0.00027}$ & ${0.08906}^{+0.00028}_{-0.00028}$ & ${0.08897}^{+0.00027}_{-0.00027}$ & ${0.08901}^{+0.00028}_{-0.00028}$ \\
$a/R_*$ \dotfill & Semi-major axis in stellar radii \dotfill & ${6.747}^{+0.094}_{-0.09}$ & ${6.797}^{+0.093}_{-0.091}$ & ${6.762}^{+0.095}_{-0.092}$ & ${6.791}^{+0.095}_{-0.093}$ \\
$T_{14}$ \dotfill & Total transit duration (days) \dotfill & ${0.12366}^{+0.00047}_{-0.00047}$ & ${0.12351}^{+0.00047}_{-0.00047}$ & ${0.10321}^{+0.00049}_{-0.00049}$ & ${0.12351}^{+0.00048}_{-0.00047}$ \\
$b$ \dotfill & Transit impact parameter \dotfill & ${0.7366}^{+0.0093}_{-0.01}$ & ${0.7295}^{+0.0098}_{-0.01}$ & ${0.7329}^{+0.0097}_{-0.010}$ & ${0.729}^{+0.01}_{-0.01}$ \\
$\sqrt{e}\cos{\omega_*}$ \dotfill & Eccentricity parameter 1  \dotfill & ${0.5188}^{+0.0015}_{-0.0014}$ & ${0.5188}^{+0.0015}_{-0.0014}$ & ${0.5188}^{+0.0015}_{-0.0015}$ & ${0.5187}^{+0.0015}_{-0.0014}$ \\
$\sqrt{e}\sin{\omega_*}$ \dotfill & Eccentricity parameter 2 \dotfill & ${-0.1016}^{+0.0082}_{-0.009}$ & ${-0.0987}^{+0.0072}_{-0.0078}$ & ${-0.0966}^{+0.0083}_{-0.0090}$ & ${-0.0962}^{+0.0078}_{-0.0084}$ \\
$u1_0$ \dotfill &  Linear LDC for \tess \dotfill& ${0.139}^{+0.037}_{-0.037}$ & ${0.159}^{+0.036}_{-0.036}$ & ${0.286}^{+0.042}_{-0.042}$ & ${0.153}^{+0.036}_{-0.037}$ \\
$u2_0$ \dotfill &  Quadratic LDC for \tess \dotfill& ${0.275}^{+0.042}_{-0.041}$ & ${0.274}^{+0.042}_{-0.042}$ & ${0.286}^{+0.042}_{-0.042}$ & ${0.282}^{+0.042}_{-0.042}$ \\
$u1_1$ \dotfill &  Linear LDC for RMs \dotfill& ${0.243}^{+0.016}_{-0.015}$ & ${0.253}^{+0.015}_{-0.014}$ & $\mathrm{...}$ & ${0.217}^{+0.017}_{-0.017}$ \\
$u2_1$ \dotfill &  Quadratic LDC for RMs \dotfill& ${0.306}^{+0.015}_{-0.015}$ & ${0.304}^{+0.012}_{-0.013}$ & $\mathrm{...}$ & ${0.301}^{+0.017}_{-0.017}$ \\
\enddata
\label{tab:combined}
\end{deluxetable*}
\section{Summary and Implications}
\label{sec:Summary}
Perhaps the most significant trend in stellar obliquity studies is that hot Jupiters --- Jupiter-mass planets, defined in our study as planets with a planet-to-star mass ratio \coolboundarywithourerr\, $\lesssim\mratio\lesssim$ \hotboundarywithourerr, on short orbital periods \citep[$a/R_*<11$;][]{Rice2021K2140,Rice2022WJs_Aligned, WangX2022WASP148, Wang2024SixWjs} --- tend to be aligned with cool host stars, whereas they are often misaligned with hot stars \citep{Winn2010}. This temperature-obliquity ($T_{\rm eff}-\lambda$) relationship is widely attributed to tidal damping mechanisms \citep{Albrecht2012, Li2016, Wang2021,  Rice2022, zanazzi2024damping}, which may be sufficiently strong to realign the obliquity of cool stars, but which are less effective around hotter stars.

The tidal realignment scenario also explains why planets with masses lower than Jupiter's --- predominantly sub-Saturns (as shown in the upper panel of Figure 2, with \mratio$<$\coolboundarywithourerr) --- consistently exhibit misalignments even around cool stars. Their low planetary masses correspond to long tidal realignment timescales, such that the systems would remain misaligned even if the host stars have thick convective envelopes and/or small radiative cores \citep{Albrecht2012, zanazzi2024damping}.

However, the tidal realignment mechanism alone cannot fully explain the trend reported in this study:
\begin{keyresult}
We find that single-star systems with high planet-to-star mass ratios (\mratio$\geq$\hotboundarywithourerr) tend to be aligned even around hot stars. This is a \siglevel$\sigma$ deviation from the stellar obliquity distribution of planets with lower mass ratios (\mratio $<$ \hotboundarywithourerr), as shown in the lower panel of Figure 2.
\end{keyresult}
Since tidal realignment is inefficient for hot stars \citep{Albrecht2012, zanazzi2024damping} --- making it unlikely that these systems are realigned through tidal interactions --- alignment for \mratio$\geq$\hotboundarywithourerr\, systems is likely primordial.

The empirical boundary between primordially aligned high-mass-ratio systems (\mratio$\geq$\hotboundarywithourerr) and often-misaligned low-mass-ratio systems (\mratio$<$\hotboundarywithourerr) suggests that stellar obliquities are more prone to excitation in systems with low planet-to-star mass ratios. Stellar obliquity measurements of compact multi-planet systems and warm Jupiters around single stars --- which are consistently aligned \citep{Albrecht2013CompactMultiAlignment, Wang2018K9, Zhou2018, Radzom2024, Rice2022WJs_Aligned, Rice2023TOI2202, Wang2024SixWjs}--- disfavor the possibility of primordial protoplanetary disk misalignments. Therefore, observed spin-orbit misalignments in systems with low planet-to-star mass ratios are likely caused by dynamical instabilities involving other planets that were originally in relatively compact configurations \citep{Rasio1996, Goldreich2004, Chatterjee2008, Nagasawa2008, Nagasawa2011, Beauge2012,  Xu2024}. The largest planet mass achievable in a compact system through core accretion is limited by the gap-opening mass \citep{Ginzburg2019}, which may align with the empirical boundary \mratio$=$ \hotboundarywithourerr\, identified between aligned high \mratio\, systems and misaligned low \mratio\, systems. Above this boundary, system would more easily become unstable, which would halt further gas accretion and prevents the formation of higher mass-ratio systems through this mechanism.

Systems with high planet-to-star mass ratios --- such as those with brown dwarf companions, or M-dwarfs hosting Jupiter-like planets --- may not form in compact configurations and would instead remain dynamically cool. Without nearby companions, dynamically isolated planets may continue to accrete gas through their circumplanetary disks even as accretion slows due to gap opening \citep{Benisty2021,Aoyama2019,Toci2020M}, which allows these planets to evolve into super-Jupiters while they maintain their primordial alignment. Alternatively, for the highest planet-to-star mass ratio objects under consideration ($M_p/M_*\geq$\hotboundarywithourerr), disk fragmentation becomes an increasingly plausible formation mechanism \citep{Xu2024arXiv241012042X}. However, the extent of the objects' migration from the outer regions of the disk \citep{Zhu2012, Galvagni2014}, where fragmentation occurs, and whether they can avoid significant perturbations and maintain low obliquities, remain uncertain.

Overall, we suggest that planets with \mratio$<$\hotboundarywithourerr\ may form in compact multi-planet systems and dynamical instabilities sculpt their properties, resulting in a spectrum of system architectures \citep{Wu2023HJsNotAlone}. These range from hyper-stable ``peas-in-a-pod" systems \citep{Weiss2018PIP, Millholland2017, Wang2017RNAAS, GoyalandWang2022} --- particularly those still locked in resonance chains \citep{Goyal2023preparation,Schmidt2024UPS,Dai2024Resonance}, which tend to be aligned \citep{Wang2018K9, Rice2023TOI2202, Dai2023TOI1136} — to meta-stable configurations, such as ultra-short-period planets \citep{Dai2018USP} and Mercury in our Solar System \citep{LithwickWu2014}, which have been found to have large inclinations. In some systems with \mratio$<$\hotboundarywithourerr\,, violent planet-planet interactions significantly increase a system's dynamical temperature, elevating orbital eccentricities and inclinations. A subset of such systems may then be dynamically cooled through tidal circularization (becoming hot Jupiters) and realignment processes, culminating in the currently observed distribution.

In contrast, systems with \mratio$\geq$\hotboundarywithourerr\, may have formed in more isolated environments. Without dynamical perturbations from compact companions, these planets could maintain their primordial alignment and continue growing on their own.


\vspace{1.5cm}

We thank Wenrui Xu, Cristobal Petrovich, and Bonan Pu for their insightful discussions. 

M.R. acknowledges support from Heising-Simons Foundation grant $\#$2023-4478 and the National Geographic Society. S.W. acknowledges support from Heising-Simons Foundation grant $\#$2023-4050. We acknowledge support from the NASA Exoplanets Research Program NNH23ZDA001N-XRP (grant $\#$80NSSC24K0153).

Funding for the \tess\, mission is provided by NASA’s Science Mission directorate. This paper includes data collected by the \tess\, mission, which are publicly available from the Mikulski Archive for Space Telescopes (MAST). 

This work has made use of data from the European Space Agency (ESA) mission
{\it Gaia} (\url{https://www.cosmos.esa.int/gaia}), processed by the {\it Gaia}
Data Processing and Analysis Consortium (DPAC,
\url{https://www.cosmos.esa.int/web/gaia/dpac/consortium}). Funding for the DPAC
has been provided by national institutions, in particular the institutions
participating in the {\it Gaia} Multilateral Agreement.

This research was supported in part by Lilly Endowment, Inc., through its support for the Indiana University Pervasive Technology Institute.

\vspace{5mm}
\facilities{WIYN/NEID, TESS (\href{https://archive.stsci.edu/doi/resolve/resolve.html?doi=10.17909/njj7-v638}{doi:10.17909/njj7-v638}) , IAPC (\href{https://catcopy.ipac.caltech.edu/dois/doi.php?id=10.26133/NEA2}{doi:10.26133/NEA2})
}

\software{\texttt{EXOFASTv2} \citep{Eastman2017, Eastman2019}, 
\texttt{iSpec} \citep{Blanco2014, Blanco2019}, 
\texttt{lightkurve} \citep{Lightkurve2018}, 
\texttt{matplotlib} \citep{hunter2007matplotlib}, 
\texttt{numpy} \citep{oliphant2006guide, walt2011numpy, harris2020array}, 
\texttt{pandas} \citep{mckinney2010data}, 
\texttt{scipy} \citep{virtanen2020scipy}, 
\texttt{SpinSpotter} \citep{Holcomb2022}.
}

\clearpage
\bibliography{main}{}
\bibliographystyle{aasjournal}

\end{document}